\documentclass[prd,aps,tightenlines,a4paper,12pt]{revtex4}

\usepackage{graphicx}
\usepackage{bm}

\usepackage{url}
\usepackage{amsmath}

\newcommand{\OO}[1]{{\mathcal O}(c^{-#1})}

\newcommand{\muas}[0]{\hbox{\rm $\mu$as}}

\newcommand{\ve}[1]{\mbox{\boldmath$#1$}}

\begin{document}

\title{Analytical solution  
for light propagation in Schwarzschild field
having an accuracy of 1 \muas}

\author{Sven \surname{Zschocke}, Sergei A. \surname{Klioner}}

\affiliation{Lohrmann Observatory, Dresden Technical University,
Mommsenstr. 13, 01062 Dresden, Germany}

\begin{abstract}
\begin{center}
{\bf GAIA-CA-TN-LO-SZ-002-2}

\medskip

issue 2, \today 
\end{center}

Numerical integration of the differential equations of light
propagation in the Schwarzschild metric shows that in some extreme
situations relevant for practical observations (e.g. for Gaia) the
well-known standard post-Newtonian formula for the boundary problem has
an error up to 16 \muas. The aim of this note is to identify the reason
for this error and to derive an extended formula accurate at the level
of 1 \muas\ as needed e.g. for Gaia.

The analytical parametrized post-post-Newtonian solution for light
propagation derived by \citet{report1} gives the solution for the
boundary problem with all analytical terms of order $\OO4$ taken into
account. Giving an analytical upper estimates of each term we
investigate which post-post-Newtonian terms may play a role for an
observer in the solar system at the level of 1 \muas. We conclude that
only one post-post-Newtonian term remains important for this numerical accuracy and derive
a simplified analytical solution for the boundary problem for light
propagation containing all the terms that are indeed relevant at the level of 1
\muas. The derived analytical solution has been verified using the
results of a high-accuracy numerical integration of differential
equations of light propagation and found to be correct at the level 
well below 1 \muas\ for arbitrary observer situated within the solar system.

\end{abstract}

\keywords{}
\pacs{}

\maketitle

\newpage

\tableofcontents

\newpage

\section{Introduction}

It is well known that adequate relativistic modelling is indispensable
for the success of microarcsecond space astrometry. One of the most important
relativistic effects for astrometric observations in the solar system
is the gravitational light deflection. The largest contribution in the
light deflection comes from the spherically symmetric (Schwarzschild)
parts of the gravitational fields of each solar system body. Although
the planned astrometric satellites Gaia, SIM, etc. will not observe very
close to the Sun, they can observe very close to the giant planets also
producing significant light deflection. This poses the problem of
modelling this light deflection with a numerical accuracy of better
than 1 \muas.

The exact differential equation of motion for a light ray in the
Schwarzschild field can be solved numerically as well as analytically.
However, the exact analytical solution is given in terms of elliptic
integrals, implying numerical efforts comparable with direct
numerical integration, so that approximate analytical solutions are
usually used. In fact, the standard parametrized post-Newtonian (PPN)
solution is sufficient in many cases and has been widely applied. So
far, there was no doubt that the post-Newtonian order of approximation
is sufficient for astrometric missions even up to microarcsecond level
of accuracy, besides astrometric observations close to the edge of the
Sun. However, a direct comparison reveals a deviation between the
standard post-Newtonian approach and the exact numerical solution of
the geodetic equations. In particular, we have found a difference of
up to 16 \muas\ in light deflection for solar system objects observed
close to giant planets. This error has triggered detailed numerical
and analytical investigations of the problem.

Usually, in the framework of general relativity or the PPN formalism
analytical orders of smallness of various terms are considered. Here
the role of small parameter is played by $c^{-1}$ where $c$ is the
light velocity. Standard post-Newtonian and post-post-Newtonian
solutions are derived by retaining terms of relevant analytical orders
of magnitude. On the other hand, for practical calculations only
numerical magnitudes of various terms are relevant. In this note we
attempt to close this gap and combine the analytical
post-post-Newtonian solution derived in \citet{report1} with
estimates of numerical magnitudes of various terms. In
this way we will derive a compact analytical solution for the boundary
problem for light propagation where all terms are indeed relevant at
the level of 1 \muas. The derived analytical solution is then verified
using high-accuracy numerical integration of the
differential equations of light propagation and found to be correct at
the level well below 1 \muas.


We use fairly standard notations: 

\begin{itemize}

\item $G$ is the Newtonian constant of gravitation;

\item $c$ is the velocity of light;

\item $\beta$ and $\gamma$ are the parameters of the Parametrized Post-Newtonian
(PPN) formalism which characterize possible deviation of the physical
reality from general relativity theory ($\beta=\gamma=1$ in general
relativity);

\item lower case Latin indices $i$, $j$, \dots take values 1,
2, 3;

\item lower case Greek indices $\mu$, $\nu$, \dots take values 0, 1,
2, 3;

\item repeated indices imply the Einstein's summation irrespective of
their positions (e.g. $a^i\,b^i=a^1\,b^1+a^2\,b^2+a^3\,b^3$ and
$a^\alpha\,b^\alpha=a^0\,b^0+a^1\,b^1+a^2\,b^2+a^3\,b^3$);

\item a dot over any quantity designates the total derivative with
respect to the coordinate time of the corresponding reference system:
e.g. $\dot a=\displaystyle{da\over dt}$;

\item the 3-dimensional coordinate quantities (``3-vectors'') referred to
the spatial axes of the corresponding reference system are set in
boldface: $\ve{a}=a^i$;

\item the absolute value (Euclidean norm) of a ``3-vector'' $\ve{a}$ is
denoted as $|\ve{a}|$ or, simply, $a$ and can be computed as
$a=|\ve{a}|=(a^1\,a^1+a^2\,a^2+a^3\,a^3)^{1/2}$;

\item the scalar product of any two ``3-vectors'' $\ve{a}$ and $\ve{b}$
with respect to the Euclidean metric $\delta_{ij}$ is denoted by
$\ve{a}\,\cdot\,\ve{b}$ and can be computed as
$\ve{a}\,\cdot\,\ve{b}=\delta_{ij}\,a^i\,b^j=a^i\,b^i$;

\item the vector product of any two ``3-vectors'' $\ve{a}$ and $\ve{b}$
is designated by $\ve{a}\times\ve{b}$ and can be computed as
$\left(\ve{a}\times\ve{b}\right)^i=\varepsilon_{ijk}\,a^j\,b^k$, where
$\varepsilon_{ijk}=(i-j)(j-k)(k-i)/2$ is the fully antisymmetric
Levi-Civita symbol;

\item for any two vectors $\ve{a}$ and $\ve{b}$ the angle between them
is designated as $\delta(\ve{a},\ve{b})$.

\end{itemize}


The paper is organized as follows. In Section
\ref{section-schwarzschild} we present the exact differential
equations. High-accuracy numerical integration of these equations is
discussion in Section \ref{Section:numerical_integration}.  In Section
\ref{section-standard-pN} we discuss the standard post-Newtonian
approximation and demonstrate the problem with the standard
post-Newtonian solution by direct comparison between numerical results
and the PPN solution. In Section \ref{section-ppN-solution}, the
formulas for the boundary problem in post-post-Newtonian approximation
are considered. A detailed estimation of all relevant terms is given,
and simplified expressions are derived. We demonstrate by explicit
numerical examples the applicability of this analytical approach for
the GAIA astrometric mission. In Section \ref{section-stars} we
consider the important case of objects situated infinitely far from
the observer as a limit of the boundary problem.  The results are
summarized in Section \ref{section-conclusion}. In the Appendices
detailed derivations for a number of analytical formulas are given.

\section{Schwarzschild metric and null geodesics in harmonic coordinates}
\label{section-schwarzschild}

For the reasons given above we need a tool to calculate the real numerical 
accuracy of some analytical formulas for the light propagation. To this end,
we consider the exact Schwarzschild metric and its null geodesics in 
harmonic gauge. Those exact differential equations for the null geodesics
will be solved numerically with high accuracy (see below) and that numerical
solution provides the required reference.

\subsection{Metric tensor}

As it has been already discussed in \citet{report1} in harmonic gauge
\begin{equation}
\label{harmonic-conditions}
\frac{\partial \left( \sqrt{- g} \, g^{\alpha \beta} \right)}{\partial x^{\beta}}= 0
\end{equation}

\noindent
the components of the covariant metric tensor of the Schwarzschild
solution are given by
\begin{eqnarray}
g_{00} &=& - \, \frac{1-a}{1+a} ,
\nonumber\\
g_{0i} &=& 0 ,
\nonumber\\
g_{ij} &=& \left(1 + a \right)^2 \, \delta_{i j} \, + \,
\frac{a^2}{x^2} \, \frac{1+a}{1-a} \, x^i \, x^j
\label{exact_5}
\end{eqnarray}

\noindent
where
\begin{equation}
a={m\over x},
\end{equation}

\noindent
$m={GM\over c^2}$ is the Schwarzschild radius of a body with mass
$M$. The contravariant components read
\begin{eqnarray}
g^{00} &=& \frac{1 + a}{1 - a} \,,
\nonumber\\
g^{0i} &=& 0 \,,
\nonumber\\
g^{ij} &=& \frac{1}{\left(1 + a \right)^2} \; \delta_{i j}
\; - \; \frac{a^2}{x^2} \; \frac{1}{\left(1 + a \right)^2}
\; x^i \, x^j \,.
\label{exact_10}
\end{eqnarray}

\noindent
Considering that the determinant of the metric can be computed as
\begin{equation}
  \label{eq:determinant}
  g=-(1+a)^4,
\end{equation}

\noindent
one can easily check that this metric satisfies the harmonic conditions
(\ref{harmonic-conditions}).

\subsection{Christoffel symbols}

The Christoffel symbols of second kind are defined as
\begin{eqnarray}
\Gamma^{\mu}_{\alpha \beta} &=& \frac{1}{2}\;g^{\mu \nu}\;
\left( \frac{\partial g_{\nu \alpha} }{\partial x^{\beta}} \;
+ \; \frac{\partial g_{\nu \beta} }{\partial x^{\alpha}} \; - \;
\frac{\partial g_{\alpha \beta} }{\partial x^{\nu}} \right) \,.
\label{christoffel}
\end{eqnarray}

\noindent
Using (\ref{exact_5}) and (\ref{exact_10}) one gets
\begin{eqnarray}
\Gamma^0_{0 i} &=& \frac{a}{x^2} \; \frac{1}{1 - a^2} \; x^i\,,
\nonumber\\
\Gamma^i_{0 0} &=& \frac{a}{x^2} \; \frac{1-a}{(1 + a)^3} \; x^i \,,
\nonumber\\
\Gamma^i_{j k} &=& \frac{a}{x^2} \; x^i \; \delta_{j k} \; - \;
\frac{a}{x^2} \; \frac{1}{1 + a} \; \left( x^j \; \delta_{i k} \; + \;
x^k \; \delta_{i j} \right)
\; - \; \frac{a^2}{x^4} \; \frac{2 - a}{1 - a^2}\;x^i\;x^j\;x^k\,,
\label{exact_15}
\end{eqnarray}

\noindent
and all other Christoffel symbols vanish.

\subsection{Isotropic condition}

As it has been pointed out in Section II.C of \citet{report1} the
condition of isotropy
\begin{eqnarray}
g_{\alpha \beta} \; \frac{d \, x^{\alpha}}{d \, \lambda} \;
\frac{d \, x^{\beta}}{d \, \lambda} &=& 0 \,,
\label{isotropic_5}
\end{eqnarray}

\noindent
leads to the following integral of the equations of light propagation
\begin{eqnarray}
s &=& \frac{1 - a}{1 + a}\;
\left( 1 - a^2 + \frac{a^2}{x^2} (\ve{x} \cdot \ve{\mu})^2 \right)^{-1/2}\,,
\label{isotropic_15}
\end{eqnarray}

\noindent
where $\mu^i$ is the coordinate direction of propagation ($\ve{\mu}
\cdot \ve{\mu} = 1$), $\ve{x}$ is the position of the photon and $s$
is the absolute value of the coordinate light velocity normalized by
$c$\,: $s=|\dot{\ve{x}}|/c$. 

\subsection{Equation of isotropic geodesics}

Reparametrizing the geodetic equations
\begin{eqnarray}
\frac{d^2 x^{\mu}}{d \lambda^2} \; + \; \Gamma^{\mu}_{\alpha \beta} \;
\frac{d x^{\alpha}}{d \lambda}\,\frac{d x^{\beta}}{d \lambda} &=& 0\,,
\label{geodetic_5}
\end{eqnarray}

\noindent
by coordinate time $t = x^0$ (see e.g. Section II.D of \citet{report1})
and using the Christoffel symbols computed above one gets the differential
equations for the light propagation in metric (\ref{exact_5}):
\begin{eqnarray}
\ddot{\ve{x}} &=&
\frac{a}{x^2} \left[ - c^2 \frac{1 - a}{(1 + a)^3} - \dot{\ve{x}} \cdot \dot{\ve{x}}
+ a \frac{2 - a}{1 - a^2} \left( \frac{ {\ve{x}} \cdot \dot{\ve{x}}}{x} \right)^2
\right] \ve{x}
 +  2 \frac{a}{x^2} \;\frac{2 - a}{1 - a^2}
( \ve{x} \cdot \dot{\ve{x}}) \, \dot{\ve{x}} \,.
\label{exact_25}
\end{eqnarray}

\noindent
Eq. (\ref{isotropic_15}) for the isotropic condition together with
$\dot{\ve{x}}\cdot\dot{\ve{x}} = c^2\,s^2$ could be used to avoid the
term containing $\dot{\ve{x}} \cdot \dot{\ve{x}}$, but it does not
simplify the equations.

\section{Numerical Integration of the equations of light propagation}
\label{Section:numerical_integration}

Our goal is to integrate Eq. (\ref{exact_25}) numerically to get a
solution for the trajectory of a light ray with an accuracy 
much higher than the goal accuracy of $1 \muas\approx 4.8\times
10^{-12}$.  For this numerical integration a simple FORTRAN 95 code using
quadrupole (128 bit) arithmetic has been written. Numerical integrator
ODEX \cite{ODEX} has been adapted to the quadrupole precision. ODEX is
an extrapolation algorithm based on the explicit midpoint rule. It has
automatic order selection, local accuracy control and dense output.
Using forth and back integration to estimate the accuracy, 
each numerical integration is automatically checked to
achieve a numerical accuracy of at least $10^{-24}$ in the components
of both position and velocity of the photon at each moment of time.

The numerical integration is first used to solve the initial value
problem for differential equations (\ref{exact_25}).
Eq. (\ref{isotropic_15}) should be used to choose the initial
conditions. The problem of light propagation has thus only 5 degrees
of freedom: 3 degrees of freedom correspond to the position of the photon and 
two other degrees of freedom correspond to the unit direction of
light propagation. The absolute value of the coordinate light velocity
can be computed from (\ref{isotropic_15}).
Fixing initial position of the photon $\ve{x}(t_0)$ and
initial direction of propagation $\ve{\mu}$ one gets the initial
velocity of the photon as function of $\ve{\mu}$ and $s$ computed for
given $\ve{\mu}$ and $\ve{x}$:
\begin{eqnarray}
\ve{x}(t_0) &=& \ve{x}_0 \,,
\nonumber\\
\dot{\ve{x}} (t_0) &=&  c\, s \,\ve{\mu} \,.
\label{num_5}
\end{eqnarray}

\noindent
The numerical integration yields the position $\ve{x}$ and velocity
$\dot{\ve{x}}$ of a photon as function of time $t$. The dense output
of ODEX allows one to obtain the position and velocity of the photon
on a selected grid of moments of time.  Eq. (\ref{isotropic_15}) holds
for any moment of time as soon as it is satisfied by the initial
conditions. Therefore, (\ref{isotropic_15}) can be also used to
estimate the accuracy of numerical integration at each moment of
integration.

For the purposes of this work we need
to have an accurate solution of two-value boundary problem. That is, a solution
of Eq. (\ref{exact_25}) with boundary conditions
\begin{eqnarray}
\ve{x} (t_0) &=& \ve{x}_0 ,
\nonumber\\
\ve{x} (t) &=& \ve{x} \, ,
\label{num_10}
\end{eqnarray}

\noindent
where $\ve{x}_0$ and $\ve{x}$ are two given constants, $t_0$ is
assumed to be fixed and $t$ is unknown and should be determined by
solving (\ref{exact_25}). Instead of using some numerical
methods to solve this boundary problem directly, we generate solutions of
a family of 
boundary problems from our solution of initial value problem
(\ref{num_5}). Each intermediate result computed by ODEX during
the integration with initial conditions (\ref{num_5})
gives us a high-accuracy solution of the corresponding
two-value boundary problem (\ref{num_10}): $t$ and $\ve{x}$ are 
just taken from the intermediate steps of our numerical integration. 

In the following discussion we will compare predictions of various
analytical models for the unit direction of light propagation
$\ve{n}(t)$ for a given moment of time $t$. The reference value for
these comparisons can be derived directly from the numerical integration as
\begin{equation}
  \label{eq:n-numerical}
  \ve{n}(t)={\dot{\ve{x}}(t)\over\left|\dot{\ve{x}}(t)\right|}.
\end{equation}

\noindent
The accuracy of this numerically computed $\ve{n}$ in our
numerical integrations is guaranteed to be of the order of $10^{-24}$ radiant
and can be considered as exact for our purposes.

\section{Standard post-Newtonian approach}
\label{section-standard-pN}

In this Section we will recall the standard post-Newtonian approach
and will compare the results for the light deflection with the
accurate numerical solution of the geodetic equations described in the
previous Section.

\subsection{Equations of post-Newtonian approach}

The well-known equations of light propagation in first post-Newtonian
approximation with PPN parameters have been discussed by many authors.
The differential equations for the light rays are given
by the post-Newtonian terms of Eq. (22) of \citet{report1}:
\begin{eqnarray}
\ddot{\ve{x}} &=&
- \, \left(c^2 + \gamma \, \dot x^k\,\dot x^k\right)\,{a\,\ve{x}\over x^2}
+ 2 \, (1 + \gamma ) \,
{a\,\dot{\ve{x}}\,(\dot x^k\,x^k)\over x^2} + {\cal O}(c^{-2}) \,.
\label{pN_15}
\end{eqnarray}

\noindent
The analytical solution of (\ref{pN_15}) can be written in the form
\begin{eqnarray}
\ve{x}(t) &=& \ve{x}_{\rm pN}+ {\cal O} (c^{- 4}) \,,
\label{pN_20}
\\
\ve{x}_{\rm pN} &=& \ve{x}_0 + c \, (t - t_0) \, \ve{\sigma} + \Delta\ve{x}(t) \,,
\label{pN_21}
\end{eqnarray}

\noindent
where
\begin{eqnarray}
\Delta \ve{x} (t) &=& - (1 + \gamma) m  \left( \ve{\sigma} \times
(\ve{x}_0 \times \ve{\sigma}) \left( \frac{1}{x - \ve{\sigma} \cdot \ve{x}}
- \frac{1}{x_0 - \ve{\sigma} \cdot \ve{x}_0} \right) + \ve{\sigma}\,
\log \frac{x + \ve{\sigma} \cdot \ve{x}}
{x_0 + \ve{\sigma} \cdot \ve{x}_0} \right).
\nonumber\\
\label{pN_25}
\end{eqnarray}

\noindent
Solution (\ref{pN_20})--(\ref{pN_25}) satisfies the following 
initial conditions:
\begin{eqnarray}
\ve{x}(t_0) &=& \ve{x}_0\,,
\nonumber\\
\lim\limits_{t \rightarrow -\infty}\,\dot{\ve{x}}(t) &=& c\,\ve{\sigma} \,.
\label{cauchy_5}
\end{eqnarray}

\noindent
From Eqs.~(\ref{pN_20})--(\ref{pN_25}) it is easy to derive the following expression
for the unit tangent vector at observer's position (note, in boundary problem
we consider $\ve{x}_{\rm pN}$ as the exact position $\ve{x}$, according to
Eq.~(\ref{pN_20})):
\begin{eqnarray}
\ve{n}_{\rm pN}
&=& \ve{k} - (1 + \gamma)\,m\,\frac{\ve{k} \times (\ve{x}_0 \times
\ve{x})}{x\,(x\, x_0 + \ve{x} \cdot \ve{x}_0)} \,,
\label{pN_30}
\end{eqnarray}

\noindent
where $\ve{R} = \ve{x} - \ve{x}_0$, $\ve{k} = \ve{R} /R$. 
By means of Eq.~(\ref{omega_5}) given below we obtain that for the angle 
$\delta(\ve{n}_{\rm pN},\ve{k})$ between
$\ve{n}_{\rm pN}$ and $\ve{k}$ one has (for $\gamma=1$)
\begin{eqnarray}
\delta(\ve{n}_{\rm pN},\ve{k})
&\le& \frac{4 m}{d} \;\frac{x_0}{x + x_0} \,,
\label{pN_26}
\end{eqnarray}

\noindent
where
\begin{eqnarray}
\ve{d} &=& \ve{k} \times (\ve{x}_0 \times \ve{k}) =\ve{k} \times (\ve{x} \times \ve{k}) \;.
\label{distance_1}
\end{eqnarray}

\noindent
In the limit of a source at infinity one gets
\begin{eqnarray}
\lim\limits_{x_0 \rightarrow \infty} \delta(\ve{n}_{\rm pN},\ve{k}) &\le& \frac{4 m}{d} \,.
\label{pN_27}
\end{eqnarray}

\subsection{Comparison between the post-Newtonian 
approximation and numerical solution}
\label{Sec:numerical-comparison}

In order to determine the accuracy of the standard post-Newtonian
approach we have to compare the post-Newtonian predictions of the
light deflection with the results of the numerical solution of
geodetic equations. Here, we compare the difference between the unit
tangent vector $\ve{n}_{\rm pN}$ defined by (\ref{pN_30}) and the
vector $\ve{n}$ calculated from the numerical integration using
(\ref{eq:n-numerical}).

\begin{figure}[!h]
\begin{center}
\includegraphics[scale=0.3,angle=-90]{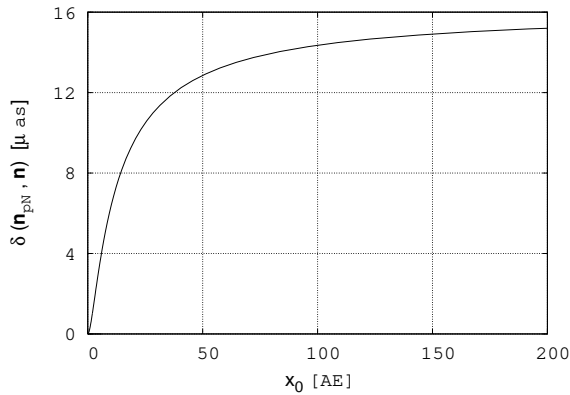}
\caption{The angle between 
$\ve{n}_{\rm pN}$ and $\ve{n}$ for Jupiter.
The vector $\ve{n}_{\rm pN}$ is evaluated by means of the
standard Newtonian formula (\ref{pN_30}), while
$\ve{n}$ is taken from the numerical integration as
described in Section \ref{Section:numerical_integration}. 
Impact parameter $d$ is taken to be the radius of Jupiter and
the distance $x$ between Jupiter and the observer is 6 AU.
}
\label{fig:numeric1}
\end{center}
\end{figure}

Having performed extensive tests, we have found that, in the real solar system,
the error of $\ve{n}_{\rm pN}$ for observations made by an observer situated in
the vicinity of the Earth attains 16 \muas. These results are
illustrated by Table~\ref{table0} and
Fig.~\ref{fig:numeric1}. Table~\ref{table0} contains the parameters we have
used in our numerical simulations as well as the maximal deviation
between $\ve{n}_{\rm pN}$ and $\ve{n}$ in each set of simulations. We
have performed simulations with different bodies of the solar systems,
assuming that the minimal impact distance $d$ is equal to the radius
of the corresponding body, and the maximal distance $x$ between the
gravitation body and the observer is given by the maximal distance
between the gravitational body and the Earth.  The simulation shows
that the error of $\ve{n}_{\rm pN}$ is generally increasing for larger
$x$ and decreasing for larger $d$.  The dependence of the error of
$\ve{n}_{\rm pN}$ for fixed $d$ and $x$ and increasing distance
between the gravitating body and the source at $x_0$ is given on
Fig.~\ref{fig:numeric1} for the case of Jupiter, $d$ being taken to be
minimal and $x$ to be maximal as given in Table~\ref{table0}.  Moreover, the
error of $\ve{n}_{\rm pN}$ is found to be proportional to $m^2$ which
leads us to the necessity to deal with the post-post-Newtonian
approximation for the light propagation.

\begin{table}[!h]
\begin{tabular}{l | c | c | c | c | c | c}
&&&&&\\[-10pt]
&\hbox to 20mm{\hfill Sun \hfill}
&\hbox to 20mm{\hfill Sun at $45^{\circ}$\hfill}
&\hbox to 20mm{\hfill Jupiter \hfill}
&\hbox to 20mm{\hfill Saturn\hfill}
&\hbox to 20mm{\hfill Uranus\hfill}
&\hbox to 20mm{\hfill Neptune\hfill} \\[3pt]
\hline
&&&&&\\[-10pt]
$m=G M/c^2$\  [m] & 1476.6 & 1476.6  & 1.40987  & 0.42215 & 0.064473 & 0.076067 \\[3pt]
$d_{\rm min}$\  [$10^6$ m] & 696.0 & 105781.7 & 71.492 & 60.268 & 25.559 & 24.764 \\[3pt]
$x_{\rm max} [{\rm AU}]$ & 1  & 1 & 6 & 11 & 21 & 31 \\[3pt]
\hline
$\delta_{\rm max} [\mu{\rm as}]$
&3187.8 & $6.32 \times 10^{-4}$ & 16.13 & 4.42 & 2.58 & 5.84 \\[3pt]
\end{tabular}
\caption{Numerical parameters of the Sun and giant planets are taken from
  \cite{Encyclopedia, IERS2003}. 
  $d_{\rm min}$ is the minimal value of the impact parameter $d$ that was used
  in the simulations.
  For each body $d_{\rm min}$ are equal
  its radius. For the Sun at $45^\circ$ the impact parameter is computed as
  $d = \sin45^\circ \times 1 \, {\rm AU}$.
  $x_{\rm max}$ is the maximal absolute value of the position of observer $x$ 
  that was used in the simulations.
  $\delta_{\rm max}$ is the maximal angle between $\ve{n}_{\rm pN}$ and $\ve{n}$ 
  found in the numerical tests.}
\label{table0}
\end{table}

\section{Post-post-Newtonian solution of boundary problem}
\label{section-ppN-solution}

In \cite{report1} an explicit analytical solution of the parametrized
post-post-Newtonian equations of light propagation in the
gravitational field of one spherically symmetric static body has been
derived. The solution $\ve{x}_{\rm ppN}$ is given by Eqs. (26)--(35)
of \cite{report1}.  Boundary problem (\ref{num_10}) has been
considered in \cite{report1}.  In this Section, we derive analytical
upper estimates of all the terms in the post-post-Newtonian solution
of the boundary problem and find which terms are responsible for numerical
errors of the post-Newtonian solution described in the previous
Section.  In this way we derive the simplest possible formulas 
that agree with exact solution at a given
numerical level.

\subsection{Analytical estimates of the individual terms in $c\,\tau$}

The propagation time between $\ve{x}_0$ and
$\ve{x}$ is given by Eq.~(50) of \cite{report1}:
\begin{eqnarray}
{\phantom{\biggr|}}_{\rm N}\biggr|\qquad &
c \, \tau =& R
\nonumber\\
{\phantom{\biggr|}}_{\rm pN}\biggr|\qquad 
&&
+ 
(1 + \gamma) \, m \, \log
\, \frac{x + x_0 + R}{x + x_0 - R}
\nonumber\\
{\phantom{\biggr|}}_{\Delta\rm pN}\biggr|\qquad 
&& + \, \frac{1}{2}\,(1 + \gamma)^2 \, m^2 \, \frac{R}{|{\ve{x}} \times {\ve{x}}_0|^2}
\, \left( (x - x_0)^2 - R^2 \right) 
\nonumber\\
{\phantom{\biggr|}}_{\rm ppN}\biggr|\qquad 
&& + \, \frac{1}{8} \, \alpha \, \epsilon \, \frac{m^2}{R} \,
\left( \frac{x_0^2 - x^2 - R^2}{x^2} \, + \,
\frac{x^2 - x_0^2 - R^2}{x_0^2} \right)
\nonumber\\
{\phantom{\biggr|}}_{\rm ppN}\biggr|\qquad 
&& + \frac{1}{4} \, \alpha \, \left(8 (1 + \gamma) - 4 \beta + 3 \epsilon\right)\,
m^2 \,
\frac{R}{|{\ve{x}} \times {\ve{x}}_0|} \;\delta (\ve{x} , \ve{x}_0)
\nonumber\\
&&
+{\cal O}(c^{- 6}) \,.
\label{iteration_7}
\end{eqnarray}

\noindent
Here and below we classify the nature of the individual terms by labels N
(Newtonian), pN (post-Newtonian), ppN (post-post-Newtonian) and
$\Delta\rm pN$ (terms that are formally of post-post-Newtonian order,
but may numarically become significantly larger than other post-post-Newtonian
terms, see below).  
Using $|\ve{x}\times\ve{x}_0|=R\,d$ where $d$ is the impact parameter
defined by (\ref{distance_1}), and assuming general-relativistic
values of all parameters $\alpha=\beta=\gamma=\epsilon=1$ 
one gets the following estimates of the sums of the terms labelled 
``ppN'' and ``$\Delta\rm pN$'':
\begin{eqnarray}
|c\,\delta\tau_{\Delta\rm pN}| 
&\le&2\,\frac{m^2}{d^2}\,R\,{4\,x\,x_0\over (x+x_0)^2}
\,\le\,2\,\frac{m^2}{d^2}\,R\,,
\label{tau_20}
\\
c\,\delta\tau=|c\,\delta\tau_{\rm ppN}| &\le& \frac{15}{4}\;\pi\;\frac{m^2}{d}\,.
\label{tau_5}
\label{tau_25}
\end{eqnarray}

\noindent
Estimates (\ref{tau_20})--(\ref{tau_5}) are proved in Appendix
\ref{appendix:c-tau1}.  Note that here are below all estimates we give
are reachable for some values of parameters and, in this sense,
cannot be improved.  From these estimates we can conclude that among
the post-post-Newtonian terms $c\,\delta\tau_{\Delta\rm pN}$ can become 
significantly larger
compared to the other post-post-Newtonian terms. 

A series of additional Monte-Carlo tests using randomly chosen
boundary conditions have been performed toverify the given estimates
of the post-post-Newtonian terms. The results of these simulations are
described in Table~\ref{table_numeric}.

The effect of $c\delta\tau$ for the Sun is less than 3.8 cm for
arbitrary boundary conditions.  Therefore, the formula for the time of
light propagation between two given points can be simplified by taking
only the relevant term:
\begin{eqnarray}
c \, \tau &=& R \, + \, (1 + \gamma)
\, m \, \log{\frac{x + x_0 + R}{x + x_0 - R}}
\nonumber
\\
&&
- \, \frac{1}{2} \, (1 + \gamma)^2
\, m^2 \, \frac{R}{|{\ve{x}} \times {\ve{x}}_0|^2}
\, \left(R^2- (x - x_0)^2 \right) +{\cal O}\left({m^2\over d}\right)+{\cal O}({m^3}).
\label{ctau-final}
\end{eqnarray}

\noindent
This expression can be written in an elegant form
\begin{eqnarray}
c \, \tau 
&=&R\, + \, (1 + \gamma) \, m \,
\log{\frac{x + x_0 + R + (1 + \gamma) \, m}{x + x_0 - R+ (1 + \gamma) \, m}}
+{\cal O}\left({m^2\over d}\right)+{\cal O}({m^3})
\label{tau_30}
\end{eqnarray}

\noindent
that has been already derived by \cite{Moyer:2000} in an inconsistent
way (see Section 8.3.1.1 and Eq.~(8-54) of \cite{Moyer:2000}). As a
criterion if the additional post-post-Newtonian term is required for a given
situation, one can use Eq. (\ref{tau_20}) giving the upper boundary
of the additional term.

\subsection{Analytical estimates of the individual terms in transformation from $\ve{k}$ to $\ve{\sigma}$}

Transformation between $\ve{k}$ and $\ve{\sigma}$ is given by
Eq. (51) of \cite{report1}:
\begin{eqnarray}
{\phantom{\biggr|}}_{\rm N}\biggr|\qquad &
{\ve{\sigma}} =& {\ve{k}} 
\nonumber\\
{\phantom{\biggr|}}_{\rm pN}\biggr|\qquad 
&& 
+ \, (1 + \gamma) \, m \, \frac{x - x_0 + R}
{|{\ve{x}} \times {\ve{x}}_0|^2} \, {\ve{k}} \times ({\ve{x}}_0 \times {\ve{x}})
\nonumber\\
{\phantom{\biggr|}}_{\Delta\rm pN}\biggr|\qquad 
&& +\frac{1}{2}\,(1 + \gamma)^2 \, \,m^2\,
{\ve{k}} \times \left( {\ve{x}}_0 \times {\ve{x}} \right)\,\frac{1}{|{\ve{x}} \times {\ve{x}}_0|^4}
\, (x + x_0) \, (x - x_0 - R) \, (x - x_0 + R)^2 
\nonumber\\
{\phantom{\biggr|}}_{\rm scaling}\biggr|\qquad 
&& 
- \, \frac{(1 + \gamma)^2}{2} \, m^2 \, \frac{(x - x_0 + R)^2}
{|{\ve{x}} \times {\ve{x}}_0|^2} \, {\ve{k}}
\nonumber\\
{\phantom{\biggr|}}_{\rm ppN}\biggr|\qquad 
&& 
+ \,m^2\,
{\ve{k}} \times \left( {\ve{x}}_0 \times {\ve{x}} \right)\,
\Bigg[ - \frac{1}{4} \, \alpha \, \epsilon \,\frac{1}{R^2}\,
\left(\frac{1}{x^2}-\frac{1}{x_0^2}\right)
\nonumber\\
{\phantom{\biggr|}}_{\rm ppN}\biggr|\qquad 
&& + \, \frac{1}{8} \, \left(8 (1 + \gamma - \alpha \, \gamma) (1 + \gamma)
- 4 \, \alpha \, \beta + 3 \alpha \, \epsilon\right)
\frac{1}{|{\ve{x}} \times {\ve{x}}_0|^3}\,
\nonumber\\
{\phantom{\biggr|}}_{\rm ppN}\biggr|\qquad 
&&\phantom{aaaaaaaaaa} \times\,
\biggl(2R^2 \left(\pi - \delta (\ve{k}, \ve{x}) \right)
+ \left(x^2 - x_0^2- R^2\right) \delta (\ve{x}, \ve{x}_0) \biggr)\Bigg]
\nonumber\\
&&
+{\cal O}(c^{- 6}) \,.
\label{k_to_sigma}
\end{eqnarray}

\noindent
Let us estimate the magnitude of the individual terms in
Eq.~(\ref{k_to_sigma}) in the angle $\delta(\ve{\sigma},\ve{k})$
between $\ve{\sigma}$ and $\ve{k}$. This angle can be computed from
vector product $\ve{k}\times\ve{\sigma}$, and, therefore, the term in
(\ref{k_to_sigma}) proportional to $\ve{k}$ and labelled as
``scaling'' plays no role.  Here and below terms proportional to
$\ve{k}$ do not influence the directions in the given order of
magnitude, but are only necessary to keep the involved vectors to have
unit length.  The total effects of the terms of the other groups on
$\ve{k}\times\ve{\sigma}$ can be estimated using
\begin{eqnarray}
\left|\, \ve{k} \times \left[ \ve{k} \times
\left( \ve{x}_0 \times \ve{x} \right) \right]  \,\right| &=&
\left|\, \ve{k} \times \left( \ve{x}_0 \times \ve{x} \right) \,\right|
\,=\, R\, d \,,
\label{sigma_10}
\end{eqnarray}

\noindent
and general-relativistic values of the parameters 
$\alpha=\beta=\gamma=\epsilon=1$
(see Appendix \ref{appendix:rho}):
\begin{eqnarray}
|\ve{\rho}_{\rm pN}| &\le&
{4m\over d}\,
\left[
\  
\begin{array}{ll}
1,\quad&  x_0\le x, \\[5pt] 
\displaystyle{\frac{x}{x + x_0}}, & x_0>x
\end{array}
\right.\,
\le {4m\over d}\,,
\label{estimate-rho-0}
\\
|\ve{\rho}_{\Delta\rm pN}| &\le& 16\, \frac{m^2}{d^3} 
\left[
\  
\begin{array}{ll}
{4\over 27}\,(x+x_0),\quad&  {1\over 2}\,x\le x_0\le x, \\[5pt] 
\displaystyle{\frac{x^2\,x_0}{(x + x_0)^2}}, & x_0<{1\over 2}\,x\ {\rm or}\ x_0>x\,,
\end{array}
\right.
\label{rho-4} 
\\
\rho=|\ve{\rho}_{\rm ppN}| &\le&  
\frac{15}{4}\,\pi\,\frac{m^2}{d^2}\,.
\label{sigma_30}
\label{sigma_40}
\end{eqnarray}

\noindent
Note that $\ve{\rho}_{\rm pN}$ and $\ve{\rho}_{\Delta\rm pN}$
themselves as well as their estimates are not continuous for
$\ve{x}\to\ve{x}_0$ since in this limit an infinitely small change of
$\ve{x}$ leads to big changes in $\ve{k}$. Discontinuity of the same
origin appears for many other terms. The limit $\ve{x}\to\ve{x}_0$ and
the corresponding discontinuity have, clearly, no physical importance.

We see that among terms of order $m^2$ only $|\ve{\rho}_{\Delta\rm
  pN}|$ cannot be estimated as ${\rm const}\times m^2/d^2$.  
The sum of the three other terms can be estimated as given by 
(\ref{sigma_30}).
The values of $\rho$ for solar system bodies are given in
Table~\ref{table2}.  In most cases these terms can be neglected at the
level of 1 \muas. Indeed, it is easy to see that $\rho$ can be comparable
with 1 \muas\ and even exceed this limit only for observations within
5 angular radii from the Sun.
Again, Monte-Carlo simulations have been performed to check the
actual maximal magnitude of these terms. The results are given in
Table~\ref{table_numeric}.  Accordingly, we obtain a simplified
formula for the transformation from $\ve{k}$ to $\ve{\sigma}$ keeping 
only the post-post-Newtonian term that
can become larger than 1 \muas\ also far from the Sun:
\begin{eqnarray}
\ve{\sigma} &=& \ve{k}\,+\,(1 + \gamma)\,m\,\frac{x - x_0 + R}
{|\ve{x} \times \ve{x}_0|^2}\,\ve{k} \times (\ve{x}_0 \times \ve{x})
\nonumber\\
&& \,+\,\frac{(1 + \gamma)^2}{2}\,m^2\,(x + x_0)\,
\frac{(x - x_0 + R)^2\,(x - x_0 - R)} {|\ve{x} \times \ve{x}_0|^4}\,
\ve{k} \times (\ve{x}_0 \times \ve{x})
+{\cal O}\left({m^2\over d^2}\right)+{\cal O}({m^3})
\,.
\nonumber\\
\label{sigma_45}
\end{eqnarray}

\noindent
This can be also written as
\begin{eqnarray}
\ve{\sigma} &=& \ve{k}\,+\ve{d}\,S\,
\left(1-S\,{1\over 2}\,(x+x_0)\left(1+{x_0-x\over R}\right)\right)
+{\cal O}\left({m^2\over d^2}\right)+{\cal O}({m^3})
\,,
\label{sigma-k-better}
\\
S&=&(1+\gamma)\,{m\over d^2}\,\left(1-{x_0-x\over R}\right)\,,
\label{sigma-k-S}
\end{eqnarray}

\noindent
where $\ve{d}$ is defined by (\ref{distance_1}).
Eq. (\ref{rho-4}) can be used as a criterion if the additional post-post-Newtonian
term in (\ref{sigma_45}) 
or (\ref{sigma-k-better}) 
is necessary for a given accuracy and configuration.

\subsection{Analytical estimates of individual terms in transformation 
from $\ve{\sigma}$ to $\ve{n}$}

Transformation between $\ve{n}$ and $\ve{\sigma}$ is given by
Eq. (55) of \cite{report1}:
\begin{eqnarray}
{\phantom{\biggr|}}_{\rm N}\biggr|\qquad &
\ve{n} =& \ve{\sigma}
\nonumber\\
{\phantom{\biggr|}}_{\rm pN}\biggr|\qquad
&&
-(1 + \gamma)\,m\,
\ve{k} \times ( \ve{x}_0 \times \ve{x} )
\frac{R}{|\, \ve{x} \times \ve{x}_0 \,|^2}\,
\left(1\,+\,\frac{\ve{k} \cdot \ve{x}}{x}\right)
\nonumber\\
{\phantom{\biggr|}}_{\rm scaling}\biggr|\qquad
&& +\frac{1}{4}\,(1 + \gamma)^2\,m^2\,
\frac{\ve{k}}{|\,\ve{x} \times \ve{x}_0\,|^2}\,
{R\over x}\,\left(1\,+\,\frac{\ve{k} \cdot \ve{x}}{x}\right)\,
(3x - x_0 - R)\,(x - x_0 + R)
\nonumber\\
{\phantom{\biggr|}}_{\Delta\rm pN}\biggr|\qquad
&& + m^2\,\ve{k} \times ( \ve{x}_0 \times \ve{x} ) \,
\Bigg[ \,(1 + \gamma)^2\,\frac{R}{|\,\ve{x} \times \ve{x}_0\,|^2}
\,\left( 1 + \frac{\ve{k} \cdot \ve{x}}{x} \right)
\frac{R\,\left(R^2-(x-x_0)^2\right)}{2\,|\,\ve{x} \times \ve{x}_0\,|^2}
\nonumber\\
{\phantom{\biggr|}}_{\rm ppN}\biggr|\qquad
&& 
\qquad
+(1 + \gamma)^2\,\frac{R}{|\,\ve{x} \times \ve{x}_0\,|^2}
\,\left( 1 + \frac{\ve{k} \cdot \ve{x}}{x} \right)\,
{1\over x}\
\nonumber\\
{\phantom{\biggr|}}_{\rm ppN}\biggr|\qquad
&& 
\qquad
-\frac{1}{2}\,\alpha\,\epsilon\,\frac{\ve{k} \cdot \ve{x}}{R\,x^4}
\nonumber\\
{\phantom{\biggr|}}_{\rm ppN}\biggr|\qquad
&&
\qquad
- {1\over 4}\,\left(8\,(1 + \gamma - \alpha\,\gamma)(1 + \gamma)
\,-4\,\alpha\,\beta\,+3\,\alpha\,\epsilon \right)
\frac{\ve{k} \cdot \ve{x}}{x^2}\,\frac{R}{|\,\ve{x} \times \ve{x}_0\,|^2}
\nonumber\\
{\phantom{\biggr|}}_{\rm ppN}\biggr|\qquad
&& 
\qquad
- {1\over 4}\,
\left(8 (1 + \gamma-\alpha\,\gamma)(1 + \gamma)\,-4\,\alpha\,\beta\,
+ 3\,\alpha\,\epsilon \right)\,\frac{R^2}{|\,\ve{x} \times \ve{x}_0\,|^3}\,
\left( \pi - \delta (\ve{k} , \ve{x}) \right) 
\, \Bigg] 
\nonumber\\
&&
+ {\cal O}(c^{- 6})\,.
\label{n_10}
\end{eqnarray}

\noindent
Let us estimate the magnitude of the individual terms in
Eq.~(\ref{n_10}) in the angle $\delta(\ve{\sigma},\ve{n})$ 
between $\ve{n}$ and $\ve{\sigma}$. This angle can be computed
from vector product $\ve{\sigma}\times\ve{n}$, and, therefore,
the term in (\ref{n_10}) 
proportional to $\ve{k}$ plays no role since $\ve{\sigma}\times\ve{k}={\cal O}(m)$.
To estimate the effects of the other terms in (\ref{n_10})
we take into account that
\begin{eqnarray}
|\,\ve{\sigma} \times \bigl( \ve{k} \times ( \ve{x}_0 \times \ve{x}) \bigr)\,| =
R\, d + {\cal O}(m),
\label{sigma-times-d}
\end{eqnarray}

\noindent
and assume again $\alpha = \beta = \gamma = \epsilon = 1$. We get
(see Appendix \ref{appendix:varphi})
\begin{eqnarray}
|\ve{\varphi}_{\rm pN}| &=& 2\,m\,
\biggl|\,\ve{\sigma} \times [ \ve{k} \times ( \ve{x}_0 \times \ve{x}) ] 
\biggr|\,
\frac{R}{|\,\ve{x} \times \ve{x}_0\,|^2}
\left(1 + \frac{\ve{k} \cdot \ve{x}}{x}\right)
 \,\le\,4\,\frac{m}{d}\,,
\label{estim_5}
\\
|\ve{\varphi}_{\Delta\rm pN}| &=& 4\,m^2\,
\biggl|\,\ve{\sigma} \times [ \ve{k} \times ( \ve{x}_0 \times \ve{x}) ] \,
\biggr|\,
\left(1 + \frac{\ve{k} \cdot \ve{x}}{x}\right)\,
\frac{R^2}{|\,\ve{x} \times \ve{x}_0\,|^4}\,\frac{R^2 - (x - x_0)^2}{2}
\nonumber\\
&&\le 4\,\frac{m^2}{d^2}\,{R\over d} \frac{4x\,x_0}{(x + x_0)^2}
\le 4\,\frac{m^2}{d^2}\,{R\over d},
\label{estim_10}
\\
\varphi&=&
|\ve{\varphi}_{\rm ppN}| \le {15\over 4}\,\pi\,{m^2\over d^2}.
\label{estim_30}
\label{n_20}
\end{eqnarray}

\noindent 
Again here $\ve{\varphi}_{ppN}$ is the sum of all the terms in (\ref{n_10})
labelled as ``ppN''.
These terms can attain 1 \muas\ only if one observes within 5 angular radii
from the Sun. The values of $\varphi$ for solar system bodies are given in
Table~\ref{table2}.  In most cases these terms can be neglected at the
level of 1 \muas. 
Again, Monte-Carlo simulations have been performed to check the
actual maximal magnitude of these terms. The results can be found in
Table~\ref{table_numeric}.  Accordingly, we obtain a simplified
formula for the transformation from $\ve{\sigma}$ to $\ve{n}$ keeping 
only the post-post-Newtonian term that can become significantly 
larger than the others:
\begin{eqnarray}
\ve{n} &=& \ve{\sigma}\,-\,(1 + \gamma)\,m\,\frac{\ve{d}}{d^2}\,
\left(1 + \frac{\ve{k} \cdot \ve{x}}{x}\right)
\nonumber\\
&& + (1 + \gamma)^2\,m^2\,\frac{\ve{d}}{d^4}\,
\left(1 + \frac{\ve{k} \cdot \ve{x}}{x}\right)\,
\frac{R^2 - (x - x_0)^2}{2\,R} 
+{\cal O}\left({m^2\over d^2}\right)+{\cal O}({m^3})
\,,
\label{n_55}
\end{eqnarray}

\noindent
where $\ve{d}$ is defined by (\ref{distance_1}). The same formula 
can be written as
\begin{eqnarray}
\ve{n} &=& \ve{\sigma}+\ve{d}\,T\,\left(1+T\,x\,
{R+x_0-x\over R+x_0+x}
\right)
+{\cal O}\left({m^2\over d^2}\right)+{\cal O}({m^3})\,,
\label{n-sigma-better}
\\
T&=& -(1 + \gamma)\,\frac{m}{d^2}\,
\left(1 + \frac{\ve{k} \cdot \ve{x}}{x}\right)\,.
\label{n-sigma-T}
\end{eqnarray}

\subsection{Analytical estimates of individual terms in transformation from
$\ve{k}$ to $\ve{n}$}

Transformation between $\ve{n}$ and $\ve{\sigma}$ is given by Eq. (56)--(57)
of \cite{report1}:
\begin{eqnarray}
{\phantom{\biggr|}}_{\rm N}\biggr|\qquad &
\ve{n} =& {\ve{k}}
\nonumber\\
{\phantom{\biggr|}}_{\rm pN}\biggr|\qquad &&
 - (1 + \gamma) \, m \,\frac{\ve{k} \times
(\ve{x}_0 \times \ve{x})}{x\left(x\,x_0 + \ve{x} \cdot \ve{x}_0\right)}
\nonumber\\
{\phantom{\biggr|}}_{\Delta\rm pN}\biggr|\qquad &&
 - (1 + \gamma) \, m \,\frac{\ve{k} \times
(\ve{x}_0 \times \ve{x})}{x\left(x\,x_0 + \ve{x} \cdot \ve{x}_0\right)}\,F
\nonumber\\
{\phantom{\biggr|}}_{\rm scaling}\biggr|\qquad 
&& - \frac{1}{8}\,(1 + \gamma)^2\,\frac{m^2}{x^2}\,\ve{k}\,
\frac{{\left((x - x_0)^2 - R^2\right)}^2}{|\ve{x} \times \ve{x}_0|^2}
\nonumber\\
{\phantom{\biggr|}}_{\rm ppN}\biggr|\qquad 
&& + \,m^2\, \ve{k} \times (\ve{x}_0 \times \ve{x})\,
\Biggl[
\,{1\over 2}\,(1 + \gamma)^2\,
\frac{R^2-(x-x_0)^2}{x^2\,|\ve{x} \times \ve{x}_0|^2}
\nonumber\\
{\phantom{\biggr|}}_{\rm ppN}\biggr|\qquad 
&&  + \, \frac{1}{4} \, \alpha \, \epsilon \, \frac{1}{R}
\left(\frac{1}{R\,x_0^2} - \frac{1}{R\,x^2}
- 2\, \frac{\ve{k} \cdot \ve{x}}{x^4}\right)
\nonumber\\
{\phantom{\biggr|}}_{\rm ppN}\biggr|\qquad 
&& - \frac{1}{4}\,\left(\, 8(1 + \gamma - \alpha \gamma) (1 + \gamma) - 4\alpha \beta
+ 3\, \alpha\, \epsilon \, \right)  \, R\,\frac{\ve{k} \cdot \ve{x}}
{x^2\,|\, \ve{x} \times \ve{x}_0 \,|^2}
\nonumber\\
{\phantom{\biggr|}}_{\rm ppN}\biggr|\qquad 
&& + \frac{1}{8}\, \left(8 (1 + \gamma - \alpha \, \gamma) (1 + \gamma)
- 4 \, \alpha \, \beta + 3 \alpha \, \epsilon\right) \,
\frac{x^2 - x_0^2 - R^2}{|{\ve{x}} \times {\ve{x}}_0|^3} \,\delta(\ve{x} , \ve{x}_0)
\Biggr]
\nonumber\\
&&
\, + \, {\cal O} \left( c^{- 6} \right)
\label{n_60}
\end{eqnarray}

\noindent
where
\begin{equation}
F=-(1 + \gamma) \,m\, \frac{x + x_0}{x\,x_0 + \ve{x} \cdot \ve{x}_0}\,.
\label{n_65}
\end{equation}

\noindent
As in other cases our goal is to
estimate the effect of the individual terms in
Eq.~(\ref{n_60}) on the angle $\delta(\ve{k},\ve{n})$ 
between $\ve{k}$ and $\ve{n}$. This angle can be computed
from vector product $\ve{k}\times\ve{n}$. 
The term in (\ref{n_65}) 
proportional to $\ve{k}$ obviously plays no role here and can be ignored.
For the other terms taking into account Eq. (\ref{sigma_10})
and considering the general-relativistic values 
$\alpha=\beta=\gamma=\epsilon=1$ one gets
(see Appendix \ref{appendix:omega})
\begin{eqnarray}
|\ve{\omega}_{\rm pN}| &=& 2\,m \,\frac{1}{x}\,
\, \frac{\left|\ve{k} \times (\ve{x}_0 \times \ve{x})\right|}
{x\,x_0 + \ve{x} \cdot \ve{x}_0}
\,\le\, 4\,\frac{m}{d}\,\frac{x_0}{x \, + \,x_0}\,\le\, 4\,\frac{m}{d}\,,
\label{omega_5}
\\
\nonumber\\
|\ve{\omega}_{\Delta\rm pN}| &=& 2\,m \,\frac{1}{x}\,
\frac{\left|\ve{k} \times (\ve{x}_0 \times \ve{x})\right|}
{x\,x_0 + \ve{x} \cdot \ve{x}_0}\,|F|
\nonumber\\
&\le& 
16\,\frac{m^2}{d^3}\,\frac{R\,x\,x_0^2}{(x+x_0)^3}
\,\le\, 16\,\frac{m^2}{d^3}\,\frac{x\,x_0^2}{(x+x_0)^2}\le 16 {m^2\over d^2}\, {x\over d}\,,
\label{omega-1}
\end{eqnarray}
\noindent
or, alternatively,
\begin{eqnarray}
|\ve{\omega}_{\Delta\rm pN}| &\le&  
{64\over 27}\,\frac{m^2}{d^2}\,{R\over d}\,.
\label{omega-1-alternative}
\end{eqnarray}
\noindent
We give four possible estimates of $|\ve{\omega}_{\Delta\rm pN}|$. 
These estimates can be useful in different situations.
Note that the last estimate in (\ref{omega-1})
and the estimate in (\ref{omega-1-alternative})
cannot be related to each other and reflect different properties of 
$|\ve{\omega}_{\Delta\rm pN}|$ as function of multiple variables.

The sum of all the terms in (\ref{n_60}) labelled as ``ppN'' is
denoted as $\ve{\omega}_{ppN}$ and can be estimated as
\begin{eqnarray}
\omega&=&|\ve{\omega}_{\rm ppN}| \le
{15\over 4}\,\pi\, {m^2\over d^2}\,.
\label{omega_30}
\label{omega_22}
\end{eqnarray}
Again these terms can attain 1 \muas\ only for observations within 5 
angular radii from the Sun.  The values of $\omega$ for solar
system bodies are given in Table~\ref{table2}. One can see that these
terms can be neglected at the level of 1 \muas\ in most cases.
The corresponding results of our Monte-Carlo simulations 
can be found in Table~\ref{table_numeric}.  Accordingly, we obtain 
a simplified
formula for the transformation from $\ve{k}$ to $\ve{n}$ keeping 
only the terms which cannot be estimated as $m^2/d^2$:
\begin{eqnarray}
\ve{n} &=& \ve{k} - (1 + \gamma) \, m \,\frac{1}{x} \,
\frac{\ve{k} \times (\ve{x}_0 \times \ve{x})}
{ x\,x_0 + \ve{x} \cdot \ve{x}_0}\,
(1 + F) 
+{\cal O}\left({m^2\over d^2}\right)+{\cal O}({m^3})
\,,
\label{n_85}
\end{eqnarray}
\noindent
where $F$ is given by (\ref{n_65}). This can be also written as
\begin{eqnarray}
\ve{n} &=& \ve{k} 
+\ve{d}\,P\,\left(1+P\,x\,{x_0+x\over R}\right)
+{\cal O}\left({m^2\over d^2}\right)+{\cal O}({m^3})\,,
\label{n_85-better}
\\
P&=&-(1+\gamma)\,{m\over d^2}\,
\left({x_0-x\over R}+{\ve{k}\cdot\ve{x}\over x}\right)\,,
\label{P-sso}
\end{eqnarray}

\noindent
where $\ve{d}$ is given by Eq. (\ref{distance_1}).  Let us also note
that the post-post-Newtonian term in (\ref{n_85}) and (\ref{n_85-better}) is maximal for
sources at infinity:
\begin{eqnarray}
|\ve{\omega}_{\Delta\rm pN}| &\le& \lim_{x_0\to\infty} |\ve{\omega}_{\Delta\rm pN}| = 
\lim_{x_0 \to \infty} (1 + \gamma)\,m\,
\frac{1}{x}\,
\frac{\left|\ve{k} \times (\ve{x}_0 \times \ve{x})\right|}
{x\,x_0 + \ve{x} \cdot \ve{x}_0}\,|F|
= (1 + \gamma)^2\,
(1-\cos\Phi)^2\,\frac{m^2}{d^2}\,{x\over d}\,,
\nonumber\\
\label{n_95}
\end{eqnarray}

\noindent
where $\Phi=\delta(\ve{x}_0,\ve{x})$ is the angle between vectors
$\ve{x}_0$ and $\ve{x}$. 
Several useful estimates of these terms are
given by (\ref{omega-1})--(\ref{omega-1-alternative}). These estimates
can be used as a criterion which allows one to decide if
the post-post-Newtonian correction is important for a particular
situation.

\begin{figure}
\begin{center}
\includegraphics[scale=0.3,angle=-90]{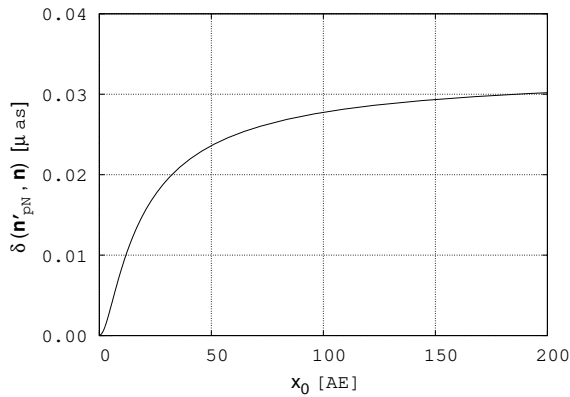}
\caption{The angle between $\ve{n}^\prime_{\rm pN}$ and $\ve{n}$ for
  Jupiter. Vector $\ve{n}^\prime_{\rm pN}$ is evaluated with the aid
  of (\ref{n_85}), while $\ve{n}$ is the high-accuracy numerical
  solution of exact geodetic equations as described in 
  Section~\ref{Section:numerical_integration}.  Impact parameter $d$ is taken
  to be the radius of Jupiter and the distance $x$ between Jupiter and
  the observer is 6 AU as on Fig. \ref{fig:numeric1}.  This figure
  demonstrates the numerical validity of (\ref{n_85-better}) at the level well below
  1 \muas.}
\label{fig:numeric2}
\end{center}
\end{figure}

Using estimate (\ref{n_95}) and the parameters of the solar system bodies
given in Table \ref{table0} one gets the maximal values of the
post-post-Newtonian correction shown in Table \ref{table5}.  For
grazing rays one can apply $\cos\Phi \simeq -1$, while for the Sun at
$45^{\circ}$ one can apply $\cos \Phi \simeq -1 / \sqrt{2}$.
Comparing these values with those in the last line of Table
\ref{table0} one sees that the post-post-Newtonian correction matches
the error of the standard post-Newtonian formula.  The deviation for a
grazing ray to the Sun is a few \muas\ and originates from the
post-post-Newtonian terms neglected in Eq.~(\ref{n_85-better}).
Vector $\ve{n}$ computed by (\ref{n_85-better})
can be denoted as $\ve{n}^\prime_{\rm pN}$ (a post-Newtonian formula
enhanced by one post-post-Newtonian term that can become large).
The numerical validity of $\ve{n}^\prime_{\rm pN}$
can be confirmed by direct comparisons of $\ve{n}^\prime_{\rm pN}$ 
and vector $\ve{n}$ computed by numerical integration
of geodetic equations as discussed in Section \ref{Sec:numerical-comparison}.
The results for Jupiter are given
on Fig.~\ref{fig:numeric2} (cf. Fig.~\ref{fig:numeric1}).

\begin{table}
\begin{tabular}{c | c | c | c | c | c | c}
&\hbox to 2mm{} Sun \hbox to 2mm{}
&\hbox to 2mm{} Sun at $45^\circ$ \hbox to 2mm{} & \hbox to 2mm{} Jupiter \hbox to 2mm{}& \hbox to 2mm{} Saturn \hbox to 2mm{} & \hbox to 2mm{} Uranus \hbox to 2mm{} & \hbox to 2mm{} Neptune \hbox to 2mm{}\\[3pt]
\hline
&&&&&\\[-10pt]
$c \,\delta \, \tau$ \ [$10^{-6}$ m] &\hbox to 2mm{} $36906.0$ \hbox to 2mm{}& $242.9$ & $0.328$ & $0.036$ & $0.002$ & $0.003$
\\[3pt]
$\rho$,
$\varphi$, $\psi$, $\omega$\ [$10^{-3}\ \mu$as] & $10937.4$ & $0.474$ & $0.945$ & $0.120$ & $0.016$ & $0.023$
\end{tabular}
\caption{
Numerical values of the analytical 
upper estimates of the post-post-Newtonian
terms of order of ${\cal O}(m^2/d)$ in Eq. (\ref{iteration_7})
of order ${\cal O}(m^2/d^2)$ in Eqs. (\ref{k_to_sigma}),
(\ref{n_10}), (\ref{n_60}), and (\ref{sigma-n-stars}).
The analytical estimates are given by Eqs. (\ref{tau_25}),
(\ref{sigma_40}), 
(\ref{n_20}),  (\ref{omega_22}), and (\ref{psi-estimate}), respectively. 
One can see that at the level of 10 cm in the distance and 1 \muas\ in angles
these terms
are irrelevant except for observations within 5 angular radii from the Sun.
}
\label{table2}
\end{table}

\begin{table}
\begin{tabular}{ c | c | c | c | c | c | c}
& \hbox to 2mm{} Sun \hbox to 2mm{}
&\hbox to 2mm{} Sun at $45^\circ$ \hbox to 2mm{} & \hbox to 2mm{} Jupiter \hbox to 2mm{}& \hbox to 2mm{} Saturn \hbox to 2mm{} & \hbox to 2mm{} Uranus \hbox to 2mm{} & \hbox to 2mm{} Neptune \hbox to 2mm{}\\[3pt]
\hline
&&&&&\\[-10pt]
$c\,\delta\,\tau$ \ [$10^{-6}$ m] &\hbox to 2mm{} $36846.9$\hbox to 2mm{}& $178.71$ & $0.327$ & $0.0348$ & $0.00192$ & $0.00274$
\\[3pt]
$\rho$ \ [$10^{-3}\times\mu$as] & $10747.6$ & $0.349$ & $0.942$ & $0.119$ & $0.0154$ & $0.0228$ \\[3pt]
$\varphi$ \ [$10^{-3}\times\mu$as] & $10713.2$ & $0.112$ & $0.942$ & $0.119$ & $0.0154$ & $0.0228$ \\[3pt]
$\psi$ \ [$10^{-3}\times\mu$as] & $10713.2$ & $0.112$ & $0.942$ & $0.119$ & $0.0154$ & $0.0228$ \\[3pt]
$\omega$ \ [$10^{-3}\times\mu$as] & $10682.8$ & $0.239$ & $0.834$ & $0.096$ & $0.0109$ & $0.0134$ 
\end{tabular}
\caption{Maximal values of the sum of the terms 
  of order of ${\cal O}(m^2/d)$ in Eq. (\ref{iteration_7}) and
  of order ${\cal O}(m^2/d^2)$ in Eqs. (\ref{k_to_sigma}),
  (\ref{n_10}), (\ref{n_60}) and (\ref{sigma-n-stars}) obtained from 
  numerical simulations. Two simulations have been performed. For the 
  first simulation $10^{8}$ starting points $\ve{x}_0$ 
  were taken within 50 AU from the relevant massive body.
  For the second simulation   
  $10^{8}$ starting points were taken at a random distance but with a 
  constraint that in each case the straight
  line between the starting and final points is tangent to the surface 
  of the body under consideration. Final point $\ve{x}$ is always 
  chosen on the orbit of the Earth.
  The position on the Earth orbit is taken randomly.
  For each of the $2\times10^8$ points the corresponding terms were evaluated
  numerically and the maximal value is given in the Table.
  The fact that these values are always below the corresponding 
  analytical estimations given in Table
  \ref{table2} can be considered 
  as additional confirmation of the estimates (\ref{tau_25}),
  (\ref{sigma_40}), 
  (\ref{n_20}),  (\ref{omega_22}), and (\ref{psi-estimate}).
  Note that the values of $\omega$ for all cases and all the values
  for ``Sun at $45^\circ$'' are systematically smaller than the estimates 
  given in  Table \ref{table2}. This behaviour is well understood and
  expected in the described set-up of the Monte-Carlo simulations.
}
\label{table_numeric}
\end{table}

\begin{table}
\begin{tabular}{c | c | c | c | c | c | c}
&\hbox to 2mm{} Sun \hbox to 2mm{} & \hbox to 2mm{}
Sun at $45^{\circ}$\hbox to 2mm{} & \hbox to 2mm{} Jupiter \hbox to 2mm{}& \hbox to 2mm{} Saturn \hbox to 2mm{} & \hbox to 2mm{} Uranus \hbox to 2mm{} & \hbox to 2mm{} Neptune \hbox to 2mm{} \\[3pt]
\hline
&&&\\[-10pt]
$\max |\ve{\omega}_{\Delta\rm pN}|$ [\muas]
&\hbox to 2mm{} $3192.8$ \hbox to 2mm{} & \hbox to 2mm{} $0.663 \times 10^{-3}$
& \hbox to 2mm{} \hbox to 2mm{} $16.11$
\hbox to 2mm{}& \hbox to 2mm{} $4.42$ \hbox to 2mm{} & \hbox to 2mm{} $2.58$
\hbox to 2mm{} & \hbox to 2mm{} $5.83$ \hbox to 2mm{} \\[3pt]
\end{tabular}
\caption{Maximal numerical values (\ref{n_95})
of the post-post-Newtonian correction 
in Eqs. (\ref{n_85}) or (\ref{n_85-better}) for the solar system bodies given in Table \ref{table0}.
A comparison of these values with the last column of Table \ref{table0}
allows one to conclude that the post-post-Newtonian term
in (\ref{n_85}) is responsible for the errors of the 
the standard post-Newtonian formula (\ref{pN_30}).}
\label{table5}
\end{table}

\section{Transformation from $\ve{k}$ to $\ve{n}$ for 
stars and quasars}
\label{section-stars}

In principle, the formulas for the boundary problem given above are valid
also for stars and quasars. However, for sufficiently large $x_0$
the formulas could be simplified. It is the purpose of this
Section to derive necessary formulas for this case.

\subsection{Transformation from $\ve{k}$ to $\ve{\sigma}$}

First, let us show that for stars and quasars the approximation
\begin{eqnarray}
\ve{\sigma} &=& \ve{k}\,
\label{stars_5}
\end{eqnarray}

\noindent
is valid for an accuracy of 1 \muas.  Using estimates
(\ref{estimate-rho-0}) and (\ref{rho-4}) for the two terms in
Eq. (\ref{sigma_45}) one can see that for $x_0\gg x$ the angle
$\delta(\ve{\sigma},\ve{k})$ can be estimated as
\begin{equation}
\delta(\ve{\sigma},\ve{k}) \le 4\, {m\over d}\,{x\over x+x_0}\,
\left(1+ 4\,{m\over d}\,{x\over d}\,{x_0\over x+x_0}\,\right).
\label{sigma-k-stars-estimates}
\end{equation}

\noindent
Clearly, $\delta(\ve{\sigma},\ve{k})$ goes to zero for $x_0\to\infty$.
Numerical values of this estimate are given in Table \ref{table_stars}
for $x_0$ equal to 1, 10 and 100 pc. Angle $\delta(\ve{\sigma},\ve{k})$ is smaller for
stars at larger distances. However, for hypothetical objects with
$x_0< 1$ pc the difference between $\ve{\sigma}$ and $\ve{k}$ must be
explicitly taken into account.

\begin{table}
\begin{tabular}{ c | c | c | c | c | c | c}
$x_0$ [pc] & \hbox to 2mm{} Sun \hbox to 2mm{}
&\hbox to 2mm{} Sun at $45^\circ$ \hbox to 2mm{}
& \hbox to 2mm{} Jupiter \hbox to 2mm{}& \hbox to 2mm{} Saturn \hbox to 2mm{} & \hbox to 2mm{} Uranus \hbox to 2mm{} & \hbox to 2mm{} Neptune \hbox to 2mm{} \\[3pt]
\hline
&&&&&\\[-10pt]
$1$ & $8.506$   & $0.056$ & $0.473$  & $0.309$ & $0.212$ & $0.382$ \\[3pt]
$10$ & $0.851$  & $5.586  \times 10^{-3}$ & $0.047$  & $0.031$ & $0.021$ & $0.038$\\[3pt]
$100$ & $0.085$ & $0.559  \times 10^{-3}$ & $4.740 \times 10^{-3}$ & $3.086 \times 10^{-3}$ &  $2.122 \times 10^{-3}$ & $ 3.819\times 10^{-3}$ \\[3pt]
\end{tabular}
\caption{Numerical values of estimate (\ref{sigma-k-stars-estimates})
in \muas\ 
for the light deflection due to the solar system bodies
with various values of $x_0$.}
\label{table_stars}
\end{table}

\subsection{Transformation from $\ve{\sigma}$ to $\ve{n}$}

As soon as we accept the equality of $\ve{\sigma}$ and $\ve{k}$ for
our case the only relevant step is  
the transformation between $\ve{\sigma}$ and $\ve{n}$. 
This transformation
in the post-post-Newtonian
approximation is given by Eqs. (53)--(54) of \cite{report1}. Introducing
impact vector computed using $\ve{\sigma}$ and the position of the
observer $\ve{x}$
\begin{equation}
\ve{d}_\sigma=\ve{\sigma}\times(\ve{x}\times\ve{\sigma})
\end{equation}

\noindent
we can re-write Eqs. (53)--(54) of \cite{report1} as
\begin{eqnarray}
{\phantom{\biggr|}}_{\rm N}\biggr|\qquad &
\ve{n}=&\ve{\sigma} 
\nonumber\\
{\phantom{\biggr|}}_{\rm pN}\biggr|\qquad 
&&
- (1+\gamma)\,m\,{\ve{d}_\sigma\over d_\sigma^2}\,
\left(1+{\ve{\sigma}\cdot\ve{x}\over x}\right)
\nonumber\\
{\phantom{\biggr|}}_{\Delta\rm pN}\biggr|\qquad 
&& 
+(1 + \gamma)^2\,m^2
\frac{\ve{d}_\sigma}{d_\sigma^3}\,{x\over d_\sigma}
{\left(1+{\ve{\sigma}\cdot\ve{x}\over x}\right)}^2 
\nonumber\\
{\phantom{\biggr|}}_{\rm scaling}\biggr|\qquad 
&& 
-\frac{1}{2}\,m^2
(1 + \gamma)^2\,\frac{\ve{\sigma}}{d_\sigma^2}
{\left(1+{\ve{\sigma}\cdot\ve{x}\over x}\right)}^2
\nonumber\\
{\phantom{\biggr|}}_{\rm ppN}\biggr|\qquad 
&&
- \frac{1}{2}\,m^2\alpha\,\epsilon\, \frac{\ve{\sigma} \cdot \ve{x}}{x^4}\,
\ve{d}_\sigma
\nonumber\\
{\phantom{\biggr|}}_{\rm ppN}\biggr|\qquad 
&&
+(1 + \gamma)^2\,m^2\frac{\ve{d}_\sigma}{d_\sigma^2}\,
{1\over x}\,\left(1+{\ve{\sigma}\cdot\ve{x}\over x}\right)
\nonumber\\
{\phantom{\biggr|}}_{\rm ppN}\biggr|\qquad 
&& 
-{1\over 4}\,\left(8\,(1+\gamma-\alpha\,\gamma)\,(1+\gamma)-4\,\alpha\,\beta+3\,\alpha\,\epsilon\right)\,m^2\,
{\ve{d}_\sigma\over d_\sigma^2}\,
{\ve{\sigma}\cdot\ve{x}\over x^2}
\nonumber\\
{\phantom{\biggr|}}_{\rm ppN}\biggr|\qquad 
&&
-{1\over 4}\,\left(8\,(1+\gamma-\alpha\,\gamma)\,(1+\gamma)-4\,\alpha\,\beta+3\,\alpha\,\epsilon\right)\,m^2\,
{\ve{d}_\sigma\over d_\sigma^3}\,
\left( \pi - \delta (\ve{\sigma} , \ve{x}) \right)
\nonumber\\
&&
+ {\cal O}(m^3),
\label{sigma-n-stars}
\end{eqnarray}

\noindent
where $d_\sigma=|\ve{d}_\sigma|=|\ve{\sigma}\times\ve{x}|$.  Now we
need to estimate the effect of the individual terms in
Eq.~(\ref{sigma-n-stars}) on the angle $\delta(\ve{\sigma},\ve{n})$ between
$\ve{\sigma}$ and $\ve{n}$. This angle can be computed from vector
product $\ve{\sigma}\times\ve{n}$.  The term in (\ref{sigma-n-stars})
proportional to $\ve{\sigma}$ obviously plays no role and can be
ignored.  For the other terms taking into account that 
$|\ve{\sigma}\times\ve{d}_\sigma|=d_\sigma$ and
considering the general-relativistic
values $\alpha=\beta=\gamma=\epsilon=1$ we get
\begin{eqnarray}
|\ve{\psi}_{\rm pN}|&=&
2\,m\,{|\ve{\sigma}\times\ve{d}_\sigma|\over d_\sigma^2}\,
\left(1+{\ve{\sigma}\cdot\ve{x}\over x}\right)
\le4\,{m\over d_\sigma}\,,
\label{psi-0}
\\
|\ve{\psi}_{\Delta\rm pN}|&=&4m^2\,
\frac{|\ve{\sigma}\times\ve{d}_\sigma|}{d_\sigma^3}\,{x\over d_\sigma}
{\left(1+{\ve{\sigma}\cdot\ve{x}\over x}\right)}^2 
\le 16\,{m^2\over d_\sigma^2}\,{x\over d_\sigma}\,,
\label{psi-3}
\\
\psi&=&|\ve{\psi}_{\rm ppN}|\le
{15\over 4}\,\pi\,{m^2\over d_\sigma^2}\,,
\label{psi-5}
\label{psi-estimate}
\end{eqnarray}

\noindent
where $\ve{\psi}_{\rm ppN}$ is the
sum of all terms of order $m^2/d_\sigma^2$ in 
(\ref{sigma-n-stars}).
Estimate (\ref{psi-estimate})
obviously agrees with estimate (\ref{n_20}) for $\varphi$. Numerical 
values of this estimate can be found in Table \ref{table2}. 
The estimates show that these terms can be neglected at the level of 
1 \muas\ except for the observations within 5 
angular radii from the Sun. Omitting these terms one gets
an expression valid at the level of 1 \muas\ in all other cases:
\begin{eqnarray}
\ve{n}&=&\ve{\sigma}+\ve{d}_\sigma\,Q\,(1+Q\,x) 
+ {\cal O}\left({m^2\over d_\sigma^2}\right)
+ {\cal O}(m^3)\,,
\label{sigma-n-stars-simplified}
\\
Q&=&- (1+\gamma)\,{m\over d_\sigma^2}\,
\left(1+{\ve{\sigma}\cdot\ve{x}\over x}\right)\,.
\label{Q-stars}
\end{eqnarray}

\noindent
Note that for $x_0\to\infty$ this coincides with
(\ref{n_85-better})--(\ref{P-sso}) and with
(\ref{n-sigma-better})--(\ref{n-sigma-T}). This formula together with
$\ve{\sigma}=\ve{k}$ can be applied for sources at distances larger
than 1 pc to attain the accuracy of 1 \muas. Alternatively
Eqs. (\ref{n_85-better})--(\ref{P-sso}) can be used for the same
purpose giving slightly better accuracy for very close stars. However,
distance information (parallax) is necessary to use
(\ref{n_85-better})--(\ref{P-sso}).

\section{Summary and concluding remarks}
\label{section-conclusion}

In this report the numerical accuracy of the post-Newtonian and
post-post-Newtonian formulas for light propagation in the parametrized
Schwarzschild field has been investigated. Analytical formulas have
been compared with high-accuracy numerical integrations of the
geodetic equations. In this way we demonstrate that the error of the
standard post-Newtonian formulas for the boundary problem (light
propagation between two given points) cannot be used at the accuracy
level of 1~\muas\ for observations performed by an observer situated
within the solar system. The error of the standard formula may attain
$\sim$ 16 \muas. Detailed analysis has shown that the error is of
post-post-Newtonian order ${\cal O}(m^2)$.  On the other hand, the
post-post-Newtonian terms are often thought to be of order $m^2/d^2$
and can be estimated to be much smaller than 1 \muas\ in this case.  
To clarify this
contradiction we have investigated the post-post-Newtonian solution
for the light propagation derived in \cite{report1}. For each
individual term in relevant formulas upper estimates have been
found. It turns out that in each case one post-post-Newtonian term may
become much larger than the other ones and cannot be estimates as
${\rm const}\times m^2/d^2$. These terms depend only on $\gamma$ and
do not come from the post-post-Newtonian terms of the corresponding
metric. The formulas for transformations between directions
$\ve{\sigma}$, $\ve{n}$ and $\ve{k}$ containing both post-Newtonian
terms and post-post-Newtonian ones that can be relevant at the level
of 10 cm for the Shapiro delay and 1 \muas\ for the directions have
been derived. The formulas are given by Eqs.  (\ref{tau_30}),
(\ref{sigma-k-better})--(\ref{sigma-k-S}),
(\ref{n-sigma-better})--(\ref{n-sigma-T}),
(\ref{n_85-better})--(\ref{P-sso}), and
(\ref{sigma-n-stars-simplified})--(\ref{Q-stars}).  These formulas
should be considered as formulas that guarantee this numerical
accuracy.

The derived analytical solution shows that no ``native''
post-post-Newtonian terms are relevant for the accuracy of 1 \muas\ in
the conditions of this note (no observations closer than five angular radii of the Sun). 
``Native'' refers here to the terms coming from the
post-post-Newtonian terms in the metric tensor. 
It is, therefore, not the post-Newtonian solution itself, but
the standard analytical way to convert the solution of the initial
value problem into the solution for the boundary problem that is
responsible for the numerical error of 16 \muas\ mentioned above.

Let us finally note that the post-post-Newtonian term in
(\ref{n_85-better})--(\ref{P-sso}) is closely related to the standard
gravitation lens formula. Here we only note that all
the formulas given in \cite{report1} and in this paper are not valid
for $d=0$ ($d$ always appear in the denominators of the relevant
formulas). On the other hand, the standard post-Newtonian lens
equation successfully treats this case, known as the Einstein ring
solution. The relation between the lens approximation
and the standard post-Newtonian expansion is a different topic which
will be considered in a subsequent paper.

\acknowledgments

This work was partially supported by the BMWi grant 50\,QG\,0601
awarded by the Deutsche Zentrum f\"ur Luft- und Raumfahrt e.V. (DLR).

\newpage

\appendix

\section{Estimates of terms in the Shapiro delay}
\label{appendix:c-tau1}

In order to get (\ref{tau_20}) we write the corresponding term as
\begin{equation}
|c\,\delta \, \tau_{\Delta\rm pN}| =
\left|\,
2 \, m^2 \, \frac{R}{|{\ve{x}} \times {\ve{x}}_0|^2}
\left((x - x_0)^2 - R^2\right)\,\right| =
2\,{m^2\over d^2}\, R\, {2z(1-\cos\Phi)\over 1+z^2-2z\,\cos\Phi}\,,
\end{equation}

\noindent
where $\Phi=\delta(\ve{x},\ve{x}_0)$ is the angle between $\ve{x}$ and
$\ve{x}_0$, and $z=x_0/x$. It is easy to see that for $0\le\Phi\le\pi$
and $z\ge0$
\begin{equation}
f_1={2z(1-\cos\Phi)\over 1+z^2-2z\,\cos\Phi}\le{4z\over(1+z)^2}\le 1.
\end{equation}

\noindent
This immediately gives (\ref{tau_20}). Here and below we always give estimates
that cannot be improved in the sense that they are reachable for certain values of the parameters.

For (\ref{tau_25}) we write

\begin{eqnarray}
c\,\delta \tau = \left|c\delta\tau_{\rm ppN}\right|
&=&
\left|\frac{1}{8} \frac{m^2}{R} \,
\left( \frac{x_0^2 - x^2 - R^2}{x^2} + 
\frac{x^2 - x_0^2 - R^2}{x_0^2} \right)
+ \frac{15}{4} 
m^2 \,
\frac{R}{|{\ve{x}} \times {\ve{x}}_0|} \;\delta (\ve{x} , \ve{x}_0)
\right|
\nonumber\\
&=& \frac{1}{4}\,\frac{m^2}{d}\;\left|\,
\sin\Phi\,\frac{z^2\,\cos\Phi - 2\,z + \cos\Phi}{1 + z^2 - 2\,z\,\cos\Phi} 
+ 15\,\Phi\,\right|\,.
\label{ctau_15} 
\end{eqnarray}

Here and below $\Phi=\delta(\ve{x},\ve{x}_0)$ is the angle between
$\ve{x}$ and $\ve{x}_0$, and $z=x_0/x$. One can show that for
$0\le\Phi\le\pi$ and $z\ge0$

\begin{eqnarray}
f_2=\left|\,
\sin \Phi\;\frac{z^2\,\cos \Phi - 2\,z + \cos \Phi}{1 + z^2 - 2\,z\,\cos \Phi} 
+ 15\,\Phi\,\right|\le 15\,\pi.
\label{f1} 
\end{eqnarray}

\noindent
and this immediately gives (\ref{tau_25}).

\section{Estimates of terms in the transformation between $\ve{\sigma}$ and $\ve{k}$}
\label{appendix:rho}

For Eq.~(\ref{estimate-rho-0}) we note that
\begin{eqnarray}
|\ve{\rho}_{\rm pN}| &=&
2\, m \, \frac{x - x_0 + R}
{|{\ve{x}} \times {\ve{x}}_0|^2} \, \left|{\ve{k}} \times ({\ve{x}}_0 \times {\ve{x}})\right|={2m\over d}\,{x - x_0 + R\over R}
\nonumber\\
&=&{2m\over d}\,
\left(
{1-z\over\sqrt{1+z^2-2z\cos\Phi}}+1
\right).
\label{for-rho-0-1}
\end{eqnarray}

\noindent
Again $\Phi=\delta(\ve{x},\ve{x}_0)$ is the angle between
$\ve{x}$ and $\ve{x}_0$, and $z=x_0/x$. One can show that for
$0\le\Phi\le\pi$ and $z\ge0$
\begin{eqnarray}
f_3={1-z\over\sqrt{1+z^2-2z\cos\Phi}}+1
&\le&
\left[
\  
\begin{array}{ll}
2,\quad&  z\le 1 \\[5pt] 
\displaystyle{\frac{2}{1 + z}}, & z>1
\end{array}
\right.\,
\le 2
\label{for-rho-0-2}
\end{eqnarray}

\noindent
and this leads to (\ref{estimate-rho-0}). The discontinuity
of $\ve{\rho}_{\rm pN}$ and its estimate at $z\to 1$ are discussed in the main text
after Eq. (\ref{rho-4}).

The term $\ve{\rho}_{\Delta\rm pN}$ can be written as
\begin{eqnarray}
|\ve{\rho}_{\Delta\rm pN}| &=& 2 \,\frac{m^2}{d^3} \, (x + x_0)\,
\left| \, \frac{(x - x_0 + R) \left[ (x - x_0)^2 - R^2 \, \right]}{R^3}
\,\right|
\nonumber\\
&=&
4\,\frac{m^2}{d^2}\,{x\over d}\,
z\,(1+z)\,{1-\cos\Phi\over 1+z^2-2z\,\cos\Phi}\,\left(1+{1-z\over\sqrt{1+z^2-2z\,\cos\Phi}}\right).
\label{for-rho-4-1}
\end{eqnarray}
\noindent
For $0\le\Phi\le\pi$ and $z\ge0$ one has
\begin{eqnarray}
f_4=z\,(1+z)\,{1-\cos\Phi\over 1+z^2-2z\,\cos\Phi}\,\left(1+{1-z\over\sqrt{1+z^2-2z\,\cos\Phi}}\right)
\le 
\left[
\  
\begin{array}{ll}
{16\over 27}\,(1+z),\ \  &  {1\over 2}\le z\le 1, \\[5pt] 
\displaystyle{4\,\frac{z}{(1+z)^2}}, & z<{1\over 2}\ {\rm or}\ z> 1\,.
\end{array}
\right.
\nonumber\\
\label{for-rho-4-2}
\end{eqnarray}
\noindent
This gives Eq.~(\ref{rho-4}). The function itself and its estimate (\ref{for-rho-4-2})
are again not continuous for $z=1$ (implying $x=x_0$). This is discussed after Eq. (\ref{rho-4}).

In order to get (\ref{sigma_40}) we write
\begin{eqnarray}
\rho=|\ve{\rho}_{\rm ppN}|&=&
m^2\,
\left|{\ve{k}} \times \left( {\ve{x}}_0 \times {\ve{x}} \right)\right|\,
\Bigg| - \frac{1}{4} \,\frac{1}{R^2}\,
\left(\frac{1}{x^2}-\frac{1}{x_0^2}\right)
+ \,\frac{15}{8}\, 
\frac{1}{|{\ve{x}} \times {\ve{x}}_0|^3}\,
\nonumber\\
&&
\phantom{aaaaaaaaaa}
\times\biggl(2R^2 \left(\pi - \delta (\ve{k}, \ve{x}) \right)
+ \left(x^2 - x_0^2- R^2\right) \delta (\ve{x}, \ve{x}_0) \biggr)
\Bigg|
\nonumber\\
&=&
{1\over 4}\,{m^2\over d^2}\,
\Bigg|
-\frac{z (z^2 - 1) \sin^3\Phi}{\left(1 + z^2 - 2\,z\,\cos\Phi\right)^2}
-15\,\arccos \frac{1 - z\,\cos \Phi}{\sqrt{1 + z^2 - 2\,z\,\cos\Phi}}
\nonumber\\
&&
\phantom{{1\over 4}\,{m^2\over d^2}\,\Bigg|}
+15\,\frac{z\left(\cos\Phi- z\right)\Phi}{1 + z^2 - 2\,z\,\cos\Phi}
+15\,\pi
\Bigg|.
\end{eqnarray}
\noindent
One can show that for $0\le\Phi\le\pi$ and $z\ge0$ 
\begin{eqnarray}
f_5&=
\Bigg|&
-\frac{z (z^2 - 1) \sin^3\Phi}{\left(1 + z^2 - 2\,z\,\cos\Phi\right)^2}
-15\,\arccos \frac{1 - z\,\cos \Phi}{\sqrt{1 + z^2 - 2\,z\,\cos\Phi}}
\nonumber\\
&& 
+15\,\frac{z\left(\cos\Phi- z\right)\Phi}{1 + z^2 - 2\,z\,\cos\Phi}
+15\,\pi
\Bigg|\le 15\pi.
\end{eqnarray}
\noindent
This immediately leads to (\ref{sigma_40}).

\section{Estimates of terms in the transformation between $\ve{n}$ and $\ve{\sigma}$}
\label{appendix:varphi}

Estimate (\ref{estim_5}) for $\ve{\varphi}_{\rm pN}$ is trivial.
For estimate (\ref{estim_10}) of $\ve{\varphi}_{\Delta\rm pN}$ we write
\begin{eqnarray}
|\ve{\varphi}_{\Delta\rm pN}| &=& 4\,m^2\,
\biggl|\,\ve{\sigma} \times [ \ve{k} \times ( \ve{x}_0 \times \ve{x}) ] \,
\biggr|\,
\left(1 + \frac{\ve{k} \cdot \ve{x}}{x}\right)\,
\frac{R^2}{|\,\ve{x} \times \ve{x}_0\,|^4}\,\frac{R^2 - (x - x_0)^2}{2}
\nonumber\\
&=&
4\,\frac{m^2}{d^2}\,{R\over d}\,
\left(1 + \frac{\ve{k} \cdot \ve{x}}{x}\right)\,
\frac{R^2 - (x - x_0)^2}{2\,R^2}
\nonumber\\
&=&
4\,\frac{m^2}{d^2}\,{R\over d}\,
\left({1-z\,\cos\Phi\over \sqrt{1+z^2-2z\,\cos\Phi}}+1\right)\,
{z\,(1-\cos\Phi)\over 1+z^2-2z\,\cos\Phi},
\end{eqnarray} 

\noindent
where again $\Phi=\delta(\ve{x},\ve{x}_0)$ is the angle between $\ve{x}$ and
$\ve{x}_0$, and $z=x_0/x$. It is easy to see that for $0\le\Phi\le\pi$
and $z\ge0$
\begin{eqnarray}
f_6=\left({1-z\,\cos\Phi\over \sqrt{1+z^2-2z\,\cos\Phi}}+1\right)\,
{z\,(1-\cos\Phi)\over 1+z^2-2z\,\cos\Phi}\le{4z\over(1+z)^2}\le 1.
\end{eqnarray} 

\noindent
This immediately leads to (\ref{estim_10}). For Eq. (\ref{estim_30}) we write

\begin{eqnarray}
\varphi=
|\ve{\varphi}_{\rm ppN}| 
&=&m^2|\ve{k} \times ( \ve{x}_0 \times \ve{x} )|\,
\Bigg|
4\,\frac{R}{|\,\ve{x} \times \ve{x}_0\,|^2}
\,\left( 1 + \frac{\ve{k} \cdot \ve{x}}{x} \right)\,
{1\over x}\
-\frac{1}{2}\,\frac{\ve{k} \cdot \ve{x}}{R\,x^4}
\nonumber\\
&&
\phantom{aaaaaa}
- {15\over 4}\,
\frac{\ve{k} \cdot \ve{x}}{x^2}\,\frac{R}{|\,\ve{x} \times \ve{x}_0\,|^2}
- {15\over 4}\,
\frac{R^2}{|\,\ve{x} \times \ve{x}_0\,|^3}\,
\left( \pi - \delta (\ve{k} , \ve{x}) \right) 
\Bigg|
\nonumber\\
&=&
{1\over 4}\,{m^2\over d^2}\,
\Bigg|
16\,{d\over x}+{d\over x}\,{\ve{k}\cdot\ve{x}\over x}-2\left({d\over x}\right)^3\,{\ve{k}\cdot\ve{x}\over x}
-15\,\left( \pi - \delta (\ve{k} , \ve{x}) \right) 
\Bigg|
\nonumber\\
&=&
{1\over 4}\,{m^2\over d^2}\,
\Bigg|
16\,\sin\Psi+\sin\Psi\,\cos\Psi-2\sin^3\Psi\,\cos\Psi-15\,\pi+15\,\Psi
\Bigg|,
\end{eqnarray} 

\noindent
where $\Psi=\delta(\ve{k},\ve{x})$ is the angle between vectors $\ve{k}$ and $\ve{x}$. Here we used
that $\ve{k}\cdot\ve{x}=x\,\cos\Psi$ and $d=|\ve{k}\times\ve{x}|=x\,\sin\Psi$.
For $0\le\Psi\le\pi$ we have
\begin{eqnarray}
f_7= \Bigg|
16\,\sin\Psi+\sin\Psi\,\cos\Psi-2\sin^3\Psi\,\cos\Psi-15\,\pi+15\,\Psi
\Bigg| \le 15\pi
\label{f7}
\end{eqnarray} 
\noindent
and this proves Eq. (\ref{estim_30}).

\section{Estimates of terms in the transformation between $\ve{n}$ and $\ve{k}$}
\label{appendix:omega}

In order to get (\ref{omega_5}) we write
\begin{eqnarray}
|\ve{\omega}_{\rm pN}| &=& 2\,m \,\frac{1}{x}\,
\, \frac{\left|\ve{k} \times (\ve{x}_0 \times \ve{x})\right|}
{x\,x_0 + \ve{x} \cdot \ve{x}_0}
=2\,{m\over d}\,{x\,x_0 - \ve{x} \cdot \ve{x}_0 \over x\,R}
=2\,{m\over d}\,{z\,(1-\cos\Phi)\over \sqrt{1+z^2-2z\,\cos\Phi}}\,.
\label{for-omega-0-1}
\end{eqnarray}

\noindent
Here again $\Phi=\delta(\ve{x},\ve{x}_0)$ is the angle between
$\ve{x}$ and $\ve{x}_0$, and $z=x_0/x$. One can show that for
$0\le\Phi\le\pi$ and $z\ge0$
\begin{eqnarray}
f_8={z\,(1-\cos\Phi)\over \sqrt{1+z^2-2z\,\cos\Phi}}\le {2z\over 1+z},
\label{for-omega-0-2}
\end{eqnarray}

\noindent
that immediately gives (\ref{omega_5}). To derive (\ref{omega-1})
and (\ref{omega-1-alternative}) we write
\begin{eqnarray}
|\ve{\omega}_{\Delta\rm pN}| &=& 2\,m \,\frac{1}{x}\,
\frac{\left|\ve{k} \times (\ve{x}_0 \times \ve{x})\right|}
{x\,x_0 + \ve{x} \cdot \ve{x}_0}\,|F|
= 4\,m^2\, 
\frac{\left|\,\ve{k} \times (\ve{x}_0 \times \ve{x})\,\right|}
{{\left(x\,x_0 + \ve{x} \cdot \ve{x}_0\right)}^2}\,
\frac{x+x_0}{x}
\nonumber\\
&=&4\,{m^2\over d^2}\,{R\over d}\,{z^2\,(1+z)\,{(1-\cos\Phi)}^2\over {(1+z^2-2z\,\cos\Phi)}^2}.
\label{for-omega-1-1}
\end{eqnarray}

\noindent
For $0\le\Phi\le\pi$ and $z\ge0$ one gets
\begin{eqnarray}
f_9={z^2\,(1+z)\,{(1-\cos\Phi)}^2\over {(1+z^2-2z\,\cos\Phi)}^2}\le{4z^2\over (1+z)^3}.
\label{for-omega-1-2}
\end{eqnarray}

\noindent
This gives the first estimate in (\ref{omega-1}).
Trivial inequalities $R\le x+x_0$, ${z^2\over(1+z)^2}\le1$ and
${z^2\over(1+z)^3}\le{4\over 27}$ give the second and third estimates in
(\ref{omega-1}) and estimate (\ref{omega-1-alternative}),
respectively. 

For (\ref{omega_22}) we write
\begin{eqnarray}
\omega=|\ve{\omega}_{\rm ppN}|&=&
m^2|\ve{k} \times ( \ve{x}_0 \times \ve{x} )|\,
\Bigg|\;
2\,
\frac{R^2-(x-x_0)^2}{x^2\,|\ve{x} \times \ve{x}_0|^2}
+ \frac{1}{4\,R}\,
\left(\frac{1}{R\,x_0^2} - \frac{1}{R\,x^2}
- 2\, \frac{\ve{k} \cdot \ve{x}}{x^4}\right)
\nonumber\\
&&
\phantom{aaa}
- \frac{15}{4}\, R\,\frac{\ve{k} \cdot \ve{x}}
{x^2\,|\, \ve{x} \times \ve{x}_0 \,|^2}
+ \frac{15}{8}\, 
\frac{x^2 - x_0^2 - R^2}{|{\ve{x}} \times {\ve{x}}_0|^3} \,\delta(\ve{x} , \ve{x}_0)
\Bigg|
\nonumber\\
&=&
{1\over 4}\,{m^2\over d^2}\,
\Bigg|
{z\,(16z-z\,\cos\Phi-15)\,\sin\Phi\over 1+z^2-2z\,\cos\Phi}
+{z(1-3z^2+2z^3\cos\Phi)\,\sin^3\Phi\over \left(1+z^2-2z\,\cos\Phi\right)^2}
\nonumber\\
&&
\phantom{aaaaaa}
+{15z\,(\cos\Phi-z)\,\Phi\over 1+z^2-2z\,\cos\Phi}
\Bigg|.
\end{eqnarray}
\noindent
For $0\le\Phi\le\pi$ and $z\ge0$ one can demonstrate that
\begin{eqnarray}
f_{10}&=&
\Bigg|
{z\,(16z-z\,\cos\Phi-15)\,\sin\Phi\over 1+z^2-2z\,\cos\Phi}
+{z(1-3z^2+2z^3\cos\Phi)\,\sin^3\Phi\over \left(1+z^2-2z\,\cos\Phi\right)^2}
+{15z\,(\cos\Phi-z)\,\Phi\over 1+z^2-2z\,\cos\Phi}
\Bigg|
\nonumber\\[10pt]
&&\quad\le 15\,\pi
\end{eqnarray}
\noindent
and this leads to (\ref{omega_22}).

\section{Estimates of terms in the transformation between $\ve{n}$ and $\ve{\sigma}$ for stars and quasars}
\label{appendix:psi}

Estimates (\ref{psi-0})--(\ref{psi-3}) are trivial. For (\ref{psi-estimate}) we write
\begin{eqnarray}
\psi=|\ve{\psi}_{\rm ppN}|&=&m^2\,|\ve{d}_\sigma|\,
\Bigg|
- \frac{1}{2}\,\frac{\ve{\sigma}\cdot\ve{x}}{x^4}
+4\frac{1}{d_\sigma^2\,x}\,
\left(1+{\ve{\sigma}\cdot\ve{x}\over x}\right)
-{15\over 4}\,
{\ve{\sigma}\cdot\ve{x}\over d_\sigma^2\,x^2}
-{15\over 4}\,
{\pi - \delta(\ve{\sigma} ,\ve{x})\over d_\sigma^3}\,
\Bigg|
\nonumber\\
&=&
{1\over 4}\,{m^2\over d_\sigma^2}\,
\Bigg|
-2\,\frac{\ve{\sigma}\cdot\ve{x}}{x}\,{d_\sigma^3\over x^3}
+\frac{d_\sigma}{x}\,{\ve{\sigma}\cdot\ve{x}\over x}
+16\,\frac{d_\sigma}{x}\,
-15\,\pi
+15\,\delta(\ve{\sigma} ,\ve{x})
\Bigg|
\nonumber\\
&=&
{1\over 4}\,{m^2\over d_\sigma^2}\,
\Bigg|
16\,\sin\Psi_\sigma
+\cos\Psi_\sigma\,\sin\Psi_\sigma
-2\,\sin^3\Psi_\sigma\,\cos\Psi_\sigma
-15\,\pi
+15\,\Psi_\sigma
\Bigg|,
\end{eqnarray}
\nonumber 
where $\Psi_\sigma=\delta(\ve{\sigma} ,\ve{x})$ is the angle
between vectors $\ve{\sigma}$ and $\ve{x}$. Here we use $d_\sigma=|\ve{\sigma}\times\ve{x}|=x\,\sin\Psi_\sigma$ 
and $\ve{\sigma}\cdot\ve{x}=x\,\cos\Psi_\sigma$. Therefore, for $0\le\Psi_\sigma\le\pi$ one can use estimate (\ref{f7})
for $f_7$ to prove (\ref{psi-estimate}).

\end{document}